\documentclass{aa}
\usepackage{graphicx}
\usepackage{lscape}
\usepackage{subfigure}
\usepackage{natbib}
\usepackage[varg]{txfonts}
\usepackage{longtable}
\usepackage{booktabs}
\usepackage[normalem]{ulem} 
\usepackage{epstopdf}
\usepackage{xcolor}
\usepackage{soul}
\usepackage{makecell}
\usepackage{multicol,tabularx,capt-of}

\usepackage{scalerel}
\usepackage{tikz}
\usetikzlibrary{svg.path}

\definecolor{orcidlogocol}{HTML}{A6CE39}
\tikzset{
  orcidlogo/.pic={
    \fill[orcidlogocol] svg{M256,128c0,70.7-57.3,128-128,128C57.3,256,0,198.7,0,128C0,57.3,57.3,0,128,0C198.7,0,256,57.3,256,128z};
    \fill[white] svg{M86.3,186.2H70.9V79.1h15.4v48.4V186.2z}
                 svg{M108.9,79.1h41.6c39.6,0,57,28.3,57,53.6c0,27.5-21.5,53.6-56.8,53.6h-41.8V79.1z M124.3,172.4h24.5c34.9,0,42.9-26.5,42.9-39.7c0-21.5-13.7-39.7-43.7-39.7h-23.7V172.4z}
                 svg{M88.7,56.8c0,5.5-4.5,10.1-10.1,10.1c-5.6,0-10.1-4.6-10.1-10.1c0-5.6,4.5-10.1,10.1-10.1C84.2,46.7,88.7,51.3,88.7,56.8z};
  }
}

\newcommand\orcidicon[1]{\href{https://orcid.org/#1}{\mbox{\scalerel*{
\begin{tikzpicture}[yscale=-1,transform shape]
\pic{orcidlogo};
\end{tikzpicture}
}{|}}}}

\usepackage{hyperref} 
\hypersetup{
    colorlinks=true,
    citecolor=blue,
    linkcolor=blue,
    urlcolor=blue,
    }
\makeatletter
\renewcommand*\aa@pageof{, page \thepage{} of \pageref*{LastPage}}
\makeatother

\bibpunct{(}{)}{;}{a}{}{,} 
\usepackage{amstext} 
\definecolor{cadmiumred}{rgb}{0.89, 0.0, 0.13}

\definecolor{ste}{rgb}{0., 0.26, 0.15}

\newcommand{\orcid}[1]{\href{https://orcid.org/#1}{\includegraphics[width=8pt]{orcid.png}}}

\newcommand{\bhm}{$M_{\rm BH}$} 
\newcommand{\snxs}{$\sigma^2_{\rm nxs}$} 
\newcommand{\nT}{$\nu_{\rm T}$}
\newcommand{\RXTE}{${\rm RXTE}$}
\newcommand{\swift}{${\it Swift}$}
\newcommand{\ASCA}{${\rm ASCA}$}

\begin{document} 
   \title{The universal shape of the X-ray variability power spectrum of AGN up to $z\sim 3$}

   \author{M. Paolillo\inst{1,2,3\orcidicon{0000-0003-4210-7693}}, 
   I. E. Papadakis\inst{4,5\orcidicon{0000-0001-6264-140X}}, 
   W.N. Brandt\inst{6,7,8\orcidicon{0000-0002-0167-2453}},
   F.E. Bauer\inst{9,10,11\orcidicon{0000-0002-8686-8737}}
   G. Lanzuisi\inst{12\orcidicon{0000-0001-9094-0984}}, 
   V. Allevato\inst{2,13\orcidicon{0000-0001-7232-5152}}, 
   O.~Shemmer\inst{14\orcidicon{0000-0003-4327-1460}},
   X. C. Zheng\inst{15\orcidicon{0000-0002-1564-0436}},
   D. De Cicco\inst{1,10\orcidicon{0000-0001-7208-5101}},
   R. Gilli\inst{12\orcidicon{0000-0001-8121-6177}},
   B. Luo\inst{16,17\orcidicon{0000-0003-2088-6403}},
   M. Thomas\inst{14\orcidicon{0000-0002-2456-3209}},
   P. Tozzi\inst{18\orcidicon{0000-0003-3096-9966}},
   F. Vito\inst{12\orcidicon{0000-0003-0680-9305}},
   Y. Q. Xue\inst{19,20\orcidicon{0000-0002-1935-8104}}
   }

   \titlerunning{Analysis of the structure function of VST-COSMOS AGN}
   \authorrunning{M. Paolillo et al.}

\institute{Dipartimento di Fisica ``Ettore Pancini'', Universit\`a di Napoli Federico II, Via Cintia, 80126, Italy    
\\e-mail: paolillo@na.infn.it
\and
INAF -- Osservatorio Astronomico di Capodimonte, Via Moiariello 16, 80131, Naples, Italy     
\and
INFN -- Unit\`a di Napoli, via Cintia 9, 80126, Napoli, Italy 
\and
Department of Physics and Institute of Theoretical and Computational Physics, University of Crete, 71003 Heraklion, Greece 
\and
Institute of Astrophysics, FORTH, GR-71110 Heraklion, Greece 
\and
Department of Astronomy and Astrophysics, 525 Davey Lab, The Pennsylvania State University, University Park, PA 16802, USA 
\and
Institute for Gravitation and the Cosmos, The Pennsylvania State University, University Park, PA 16802, USA 
\and
Department of Physics, The Pennsylvania State University, University Park, PA 16802, USA 
\and
Instituto de Astrof{\'{\i}}sica and Centro de Astroingenier{\'{\i}}a, Facultad de F{\'{i}}sica, Pontificia Universidad Cat{\'{o}}lica de Chile, Casilla 306, Santiago 22, Chile	
\and
Millennium Institute of Astrophysics, Nuncio Monse{\~{n}}or S{\'{o}}tero Sanz 100, Of 104, Providencia, Santiago, Chile 
\and
Space Science Institute, 4750 Walnut Street, Suite 205, Boulder, Colorado 80301 
\and
INAF -- Osservatorio di Astrofisica e Scienza dello Spazio di Bologna, I-40129 Bologna, Italy 
\and
Scuola Normale Superiore, Piazza dei Cavalieri 7, I-56126 Pisa, Italy 
\and
Department of Physics, University of North Texas, Denton, TX 76203, USA 
\and
Leiden Observatory, Leiden University, PO Box 9513, 2300 Leiden, RA, The Netherlands 
\and
School of Astronomy and Space Science, Nanjing University, Nanjing, Jiangsu 210093, China 
\and
Key Laboratory of Modern Astronomy and Astrophysics (Nanjing University), Ministry of Education, Nanjing 210093, China 
\and
INAF -- Osservatorio Astrofisico di Arcetri, Largo Enrico Fermi 5, Firenze, Italy 
\and
CAS Key Laboratory for Research in Galaxies and Cosmology, Department of Astronomy, University of Science and Technology of China, Hefei 230026, China 
\and
School of Astronomy and Space Sciences, University of Science and Technology of China, Hefei 230026, China\\ 
}

   \date{}
  \abstract
   {}
   {We study the ensemble X-ray variability properties of Active Galactic Nuclei (AGN) over a large range of timescales (20 ks $\leq T\leq$ 14 yrs), redshift ($0\leq z \lesssim 3$), luminosities ($10^{40} \mbox{erg s}^{-1}\leq L_X\leq 10^{46} \mbox{erg s}^{-1}$) and black hole (BH) masses ($10^6 \leq \mbox{M}_{\odot} \leq 10^9$).}
   {We propose the use of the variance–frequency diagram, as a viable alternative to the study of the power spectral density (PSD), which is not yet accessible for distant, faint and/or sparsely sampled AGN.}
   {We show that the data collected from archival observations and previous literature studies are fully consistent with a universal PSD form which does not show any evidence for systematic evolution of shape or amplitude with redshift or luminosity, even if there may be differences between individual AGN at a given redshift or luminosity. We find new evidence that the PSD bend frequency depends on BH mass and, possibly, on accretion rate. We finally discuss the implications for current and future AGN population and cosmological studies.}
   {}

   
   \maketitle



\section{Introduction}
\label{sec:Intro}
Flux variability is a defining characteristic of Active Galactic Nuclei (AGN). AGN vary on all timescales, and across the whole electromagnetic spectrum. The fastest and largest amplitude variations are observed at the  highest energies (X-rays and $\gamma$-rays), strongly suggesting that such radiation is mainly generated in small-size regions, close to the central engine. 

One of the most frequently used tools to study the observed variations is the power spectral density function (or power-spectrum for simplicity; PSD). Early studies of the X--ray variability of AGN with {\it EXOSAT} showed that the PSD has a power-law shape with a slope of $\sim -1.5$, and an amplitude that scales with the source luminosity \citep{Green93, Lawrence93}. Long \RXTE observing campaigns over many years combined with shorter XMM-{\it Newton} observations (mainly) have allowed the detailed study of AGN X--ray PSDs over a large frequency range, revealing at least one, and in some cases two, breaks in the PSD of nearby AGN \citep[e.g.][]{Uttley02, Papadakis02, Markowitz03, McHardy04, Gonzalez-Martin2012}. These should represent characteristic timescales linked to the physical process producing the observed emission.  

Most of our knowledge about AGN power-spectra in the X-ray band is derived from extensive observations of nearby and mostly low-luminosity AGN. It is not possible (yet) to estimate the PSD of AGN at larger redshifts because the available lightcurves have few points and are sparsely sampled. For this reason, 
our knowledge of the variability properties of the overall AGN population is mainly based on studies of the excess variance as a function of redshift and luminosity, using lightcurves from large samples of X-ray detected AGN in various surveys \citep[e.g][]{ Paolillo04, Papadakis08, Young12, Shemmer14, Lanzuisi14, Yang2016, Middei17, Zheng2017, Ding2018, Thomas2021}. 

Recently, \citet[][P17 hereafter]{Paolillo17} studied the X-ray variability properties of distant AGN in the Chandra Deep Field-South region (CDF-S) over 17 years. They used the normalized excess variance \snxs\
 (i.e. the average lightcurve variance corrected for the noise, see eq. 1 in P17) as a measure of the X-ray variability amplitude of the sources, and they studied the dependence of \snxs\ on X-ray luminosity in various redshift bins. They assumed power-spectrum models based on PSD analysis of nearby, bright X--ray Seyferts, and they found that the variability properties of high-$z$ AGN are consistent with a PSD described by a bending power-law, where the bend frequency (and perhaps the PSD amplitude as well) depends on the accretion rate, expressed in terms of the Eddington ratio $\lambda _E=\dot M/\dot M_{Edd}$, where $\dot M$ is the mass accretion rate and $\dot M_{Edd}$ is the Eddington accretion rate. 

In this work we expand this study, collecting several complementary AGN samples with available excess variance measurements, including the CDF-S (P17), COSMOS \citep{Lanzuisi14}, CAIXA \citep{Ponti12}, TARTARUS \citep{Oneill05} and \RXTE\ \citep{Zhang11},  in order to cover a wide range of timescales, redshifts, luminosities and black-hole (BH) masses. We also estimated the excess variance of numerous additional local AGN, using \swift/XRT and  \RXTE\ lightcurves, in order to cover timescales in-between the shortest and the longest ones probed by the \snxs\ data from the literature. 

Our objective is to study the PSD itself using excess variance measurements. In most previous works, \snxs\, has been used to investigate the dependence of AGN variability amplitude on X--ray luminosity, BH mass and redshift. 
However the dependence of \snxs\ on total lightcurve duration $T$ itself is also important, 
because the excess variance is (approximately) equal to the integral of the intrinsic power spectrum in the range of frequencies $1/T\leq \nu\leq 1/(2\Delta t_{\rm min})$, where $\Delta t_{\rm min}$ is the minimum time difference between successive points in the lightcurve\footnote{Although the excess variance is a biased estimate of the PSD integral, in the case of red-noise PSDs and sparsely sampled lightcurves, it is possible to correct \snxs\, for this effect as discussed in section 5 of \cite{Allevato13}.}  (see \S \ref{sec:vfptheory}).
Due to the close relation between the PSD and \snxs, we can compute \snxs\ from lightcurves with different duration $T$, and then plot \snxs\ as a function of \nT($\equiv 1/T$).\footnote{The additional  dependence on $1/\Delta t_{\rm min}$ is less relevant due to the steep slope of the PSD at high frequencies, and it is anyway fully taken into account in both our simulations, modeling and fitting, as we discuss in detail in the following sections.} 
We refer to the \snxs\ vs. \nT\ plot as the ``variance--frequency'' plot (VFP). The VFP provides 
information closely related to the PSD, with the advantage that it can be directly derived for large samples of faint and/or distant AGN, and thus can be studied with the aim of constraining the intrinsic PSD properties.

Although \snxs\ is easy to compute, and hence we can create a VFP even from sparsely sampled lightcurves (which are not sufficient to measure the PSD), the use of the VFP is difficult on an \textit{individual} AGN basis. In fact, given the statistical properties of the \snxs\, in the case of a single object we would need many lightcurves in order to estimate, reliably, the intrinsic variance on various timescales \citep[see][and references therein]{Allevato13}. 

On the other hand, we could consider samples of AGN which have been monitored in the same way (i.e. where $T$ and $\Delta t_{\rm min}$ are the same for all sources) to compute the mean excess variance, and use it to create the VFP. However, in the case where we use lightcurves of many AGN, one has to consider the dependence of \snxs\ on BH mass (\bhm) as well: for a given lightcurve duration, the variance decreases with increasing BH mass \citep[e.g.][]{Papadakis04, Oneill05, Ponti12}. Therefore, 
assuming we know the BH masses for all AGN in the sample, we must first
model the excess variance dependence on \bhm\ to create VFP for AGN at a fixed mass.

Following this approach, the questions we aim to investigate are the following:  a) is the measured VFP consistent with the hypothesis that the X--ray PSD has the same form in all AGN (i.e., the hypothesis that the X--ray variability mechanism is the same in all of them), and b) if yes, what are the characteristics of this ``universal'' X--ray PSD of AGN?



 The paper is organized as follows: in \S \ref{sec:vfptheory} we explain the relation between the VFP and the PSD, and how we can measure the VFP for sparsely sampled, low S/N AGN. In \S \ref{sec:highzAGN} we present the BH mass and variability measurements for high redshift sources in CDF-S and COSMOS samples, and the best-fit results to their \snxs\, -- \bhm\ relations. In \S \ref{sec:lowzAGN} we present the best-fit results for low redshift sources from literature or archival data. In \S \ref{sec:observedVFP} and \ref{sec:fitting} we present the VFP of AGN up to redshift $\sim 3$, the method we use to fit the observed VFP and the best-fit results. Finally, in \S \ref{sec:discussion}, we summarize our work and we discuss the implications of our study. 

Throughout the paper we adopt values of $H_0 = 70$ km s$^{-1}$ Mpc$^{-1}$, $\Omega_M = 0.3$, and $\Omega_\Lambda = 0.7$. 

\section{The Variance-Frequency plot: a substitute for the PSD}
\label{sec:vfptheory}

\subsection{The PSD vs Variance-Frequency plot}
Let us assume that the AGN X-ray PSD follows a relation of the form:
\begin{equation}
\label{eq:PSDmod}
\mbox{PSD}(\nu)=A\nu^{-1}\left [ 1+\left(\frac{\nu}{\nu_b}\right)^{s} \right ] ^{-1},
\end{equation}

\noindent where $A$ is the PSD normalization ($A= \mbox{PSD}(\nu_b) \times 2 \nu_b$), and $\nu_b$ is the bend frequency\footnote{The `bend' frequency is equivalent to the `break' frequency used in earlier works in the literature where two distinct power-laws were used to fit the PSD instead of a smooth function as adopted here.}. The PSD thus defined has a logarithmic slope of $-1$ at low frequencies ($\nu<<\nu_b$), which steepens to $-(1+s)$ at higher frequencies ($\nu>>\nu_b$). This model PSD is based on the results from PSD studies of nearby, low-luminosity (but X-ray bright) AGN \citep[e.g.][]{McHardy04}. Let us also assume that $\nu_b$ depends on the BH mass as follows \citep[e.g.][]{McHardy06, Gonzalez-Martin2012}:
\begin{equation}
\label{eq:nubmod}
\nu_b=B\left(\frac{M_{\rm BH}}{10^8 {\rm M}_{\odot}}\right)^{-1},
\end{equation}

\noindent where $B$ is a constant. According to Eq. (4) in \citealt{Allevato13}, the excess variance of a lightcurve of duration $T$ will be a measure of the normalized ``band'' variance, defined as follows\footnote{Note that, for simplicity, here we use $\sigma^2$ instead of $\sigma^2_{\rm band,norm}$ in \citealt{Allevato13}.}:
\begin{equation}
\label{eq:svsnumod}
\begin{aligned}
\sigma^2 (\nu_T,\nu_{max})={} & \int^{\nu_{max}}_{\nu_{T}}\mbox{PSD}(\nu)\ d\nu=\\
	        ={} & A\left [ \ln\left(\frac{\nu_{max}}{\nu_{T}}\right )-\frac{1}{s}\ln\left(\frac{\nu_b^s+\nu_{max}^s}{\nu_b^s+\nu_{T}^s}\right ) \right ], \\
\end{aligned}
\end{equation}
\noindent
where \nT=$1/T_{\rm max}$ and $\nu_{max}=1/(2\Delta t_{min}$). In the case of equally sampled light curves, with a few missing points (like the \RXTE, \swift, \ASCA\ and XMM-{\it Newton} light curves we use - see below) the shortest frequency sampled by the data is well defined, with $\Delta t_{min}=2\Delta t$, where $\Delta t$ is the bin size of the light curves. However, in the case of the unevenly sampled light curves (like CDF-S and COSMOS), $\Delta t_{min}$ is not so obvious (see for example the discussion in Appendix D in \citealp{Scargle1982}). For these light curves, we decided to accept as $\Delta t_{min}$ the shortest time difference between successive observations. For a fixed $\Delta t_{min}$, we expect a negative correlation between $\sigma^2$ and \nT, i.e. the variability amplitude should increase with increasing lightcurve duration in AGN (due to the red noise nature of the AGN PSD). 

\begin{figure}
   \centering
   \includegraphics[width=0.5\textwidth]{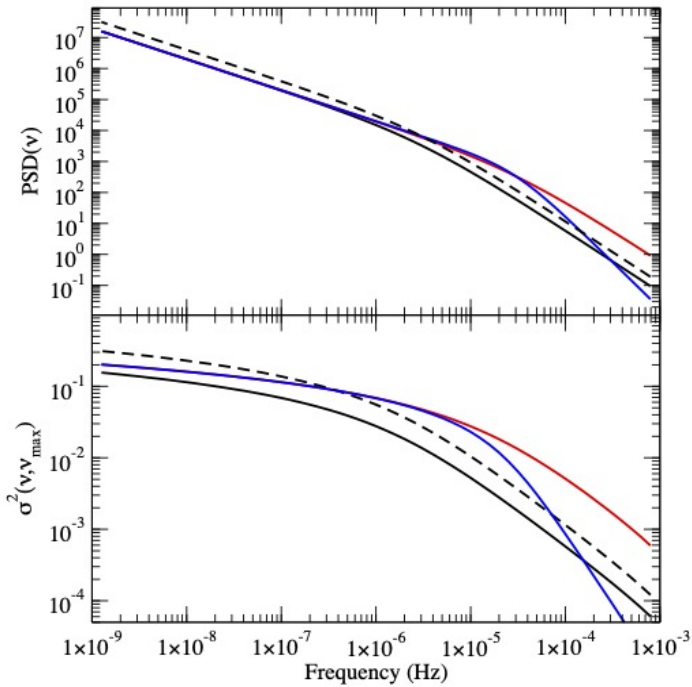}
      \caption{The PSD and the VFP (upper and lower panels, respectively) of an AGN with a BH mass of 10$^8$ M$_{\odot}$. The PSD model parameters are: $A=0.02$ Hz$^{-1}$, s=1, and $\nu_b=3\cdot 10^{-6}$ Hz (black solid lines). Black dashed lines indicate the PSD and VFP when when we increase the PSD amplitude by a factor of 2. Red lines show the PSD and VFP when we increase the bend frequency (by a factor of 10), while the blue lines show the changes when we increase the high frequency PSD slope from -2 to -3  (all the other parameters in this case are like those of the red lines). For the computation of the variance, we assumed $\nu_{max}=(1/250$ s).}
         \label{fig:vfp}
\end{figure}

Equation \ref{eq:svsnumod} shows that $\sigma^2(\nu_{T},\nu_{max})$ depends on the  shape (and normalization) of the PSD. Consequently, a plot of $\sigma^2$ as a function of \nT\ holds similar information to the PSD. We call the plot of $\sigma^2 (\nu_T,\nu_{max})$ versus \nT, the {\it Variance-Frequency Plot} (VFP) of an AGN.\footnote{We choose to define the VFP as the plot of variance vs. $1/T$, instead of $T$, so that its shape will be analogous to the PSD shape.} 

Figure~\ref{fig:vfp} shows the PSD and the VFP plot (upper and lower panels, respectively)
for an AGN with a BH mass of $10^8$ M$_{\odot}$. The PSD is computed using Eq.~(\ref{eq:PSDmod}), 
while $\sigma^2(\nu_T,\nu_{max})$ is computed using Eq.~(\ref{eq:svsnumod}), for various PSD parameters. The variance plotted in this figure should be equal to the (intrinsic) variance  of lightcurve segments with duration of $T=1/\nu$ and $\Delta t=250$ s, so that $\nu_{max}=4\cdot10^{-3}$ Hz. 

The figure shows that all the major features of the PSD are apparent in the VFP plot as well. For example, both the PSD and VFP have a power-law like shape at high frequencies, with the VFP being flatter than the PSD (note the different scale of the y-axis of the two panels in Fig.~\ref{fig:vfp}). At frequencies lower than $\nu_b$, both the PSD and VFP flatten, to a slope of $-1$ in the case of the PSD, while $\sigma^2\propto -\ln($\nT$)$, below $\nu_b$. The various lines in the same figure also show that when varying the PSD parameters, the PSD and the VFP shapes also vary in similar ways. 

It is always better to estimate the power-spectrum itself, even from a statistical point of view. The statistical properties of the periodogram (the estimator of the PSD) are far superior to the statistical properties of the excess variance as a measure of the intrinsic variance of a single source (see \citealt{Allevato13} for the statistical properties of the latter). However, as we already mentioned in \S \ref{sec:Intro}, we cannot use the available lightcurves of the high-$z$ AGN to estimate their PSD. On the other hand, we can measure their excess variance on different timescales (i.e., different \nT), hence we can compute the VFP and, as Fig.~\ref{fig:vfp} shows, infer the intrinsic PSD of the sources. 

\subsection{Measuring the VFP of active galaxies}

Ideally, we would need long and short, well sampled lightcurves of an AGN to reliably measure variance on a large range of timescales, and construct the VFP. But that is not possible with the currently available data, specially for high$-z$ sources. The available lightcurves are not good enough (either because they are too sparse and/or their signal-to-noise ratio is too low) to measure the VFP of a \textit{single} AGN. 
For that reason, we will follow a different approach to construct the VFP of active galaxies, as we describe below. 

We can consider samples of AGN, with known BH mass, and lightcurves  with the same duration ($T_{\rm obs}$) and bin size $(\Delta t$), for all AGN in each sample. We can use the lightcurves to measure the excess variance \snxs\ for each AGN in the sample. One way to study the VFP would be to choose objects with the same BH mass in each sample, and then plot \snxs\ versus $1/T_{\rm obs}$ for all of them. However, since the excess variance measurements are not Gaussian distributed and their error is unknown (see \citealp{Allevato13}), we will not be able to fit the resulting VFP and compare it with model predictions. In order to overcome this serious problem, we have to average, somehow, the measured exsccess variances.

On the other hand, we cannot simply compute the mean excess variance of all the objects in the sample, if they host different mass BHs, since the excess variance will depend on the BH mass. However, we can take advantage of this property, and produce a \snxs\ -- \bhm\ plot for all sources in each sample. The excess variance is a measure of the band variance, $\sigma^2(\nu_T,\nu_{max})$, defined by Eq.\,(\ref{eq:svsnumod}). This equation, together with Eq.\,\ (\ref{eq:nubmod}), shows how \snxs\ should depend on BH mass, depending on the sampled frequencies: if neither $\nu_{max}$ nor \nT\ are much smaller than $\nu_b$, \snxs\ should decrease with increasing BH mass, roughly in a linear way, in log-log space; if instead $\nu_{max}$ and \nT$<<\nu_b$, \snxs\ will not depend on BH mass (it will depend on \nT\ and $\nu_{\rm max}$, only; see Eq.\,(\ref{eq:svsnumod})) and the \snxs\ versus \bhm\ plot will be flat. This behavior is illustrated by the models plotted in Figs. \ref{fig:varvsbhmcdfs}--\ref{fig:varvsbhmSwiftRXTE} as dotted lines, as discussed further in \S \ref{sec:highzAGN} and \S \ref{sec:lowzAGN}.

We can fit the \snxs\ versus \bhm\ plot with a linear function of the form: 
\begin{equation} 
 \log(\sigma^2_{\rm nxs}) = \alpha(T_{\rm obs}) +\beta(T_{\rm obs})\cdot {\rm \log(M_{BH}/\rm \bar{M})}, 
 \label{eq:linemod}
\end{equation}
\noindent
where $\rm \bar{M}$ is the mean BH mass of the sample, and $\alpha(T_{\rm obs})$ is the variance of an AGN with a mass of $\bar{M}$, computed using a lightcurve of duration $T_{\rm obs}$. The key point here is that, since $\alpha(T_{\rm obs})$ is computed by fitting all the data in the \snxs\ versus \bhm\ plot, its distribution will be much closer to the Gaussian distribution, and its error will be known (from the fitting procedure). 


Thus, we can consider AGN (with known BH mass) that have been observed 
by various satellites, over different timescales, $T_{\rm obs}$,
to compute \snxs, plot \snxs\ versus \bhm, and fit the data with Eq.\,(\ref{eq:linemod}). If $\rm \bar{M}$ is the same for all samples, then the plot of $\alpha(T_{\rm obs})$ versus $1/T_{\rm obs}$ will be representative of the VFP plot for the AGN with a mass of $\rm \bar{M}$.
In this way, we can also take advantage of the study of the $\beta(T_{\rm obs})$ versus $1/T_{\rm obs}$ plot as well. 
As we showed in \S \ref{sec:vfptheory}, for a given \bhm\ (and hence $\nu_b$), the VFP shape, and thus $\beta(T_{\rm obs})$ should vary with $T_{\rm obs}$.
The way it varies depends on the relation between $\nu_b$ and \bhm. In other words, the study  of the $\beta(T_{\rm obs})$ versus $1/T_{\rm obs}$ plot will constrain the parameter $B$ in Eq.\,(\ref{eq:nubmod}). 

We plan to follow the approach we outlined above in order to study the VFP of AGN, as we describe in detail in the following sections. We will use data from various X-ray surveys for high$-z$ objects, as well as lightcurves from pointed observations of nearby objects, in order to construct the VFP of AGN, both at high and low redshift. In this way, we will be able to directly compare the low and high$-z$ objects, and investigate whether their PSDs are the same or not.

\section{The  variance -- BH mass relation of high--redshift AGN}
\label{sec:highzAGN}

\subsection{Black-hole mass measurements for the CDF-S sources}
\label{sec:cdfsbhms}

The first AGN sample we considered is derived from the CDF-S X-ray catalog of \citet{Luo17}, and the CDF-S variability measurements used in this work are described in P17; we refer to those works for specific details.
BH mass measurements for CDF-S sources are  primarily based on the measurements published by \citet{Suh2015}. These are derived from optical/near-infrared spectroscopic measurements of H$\alpha$, H$\beta$ and Mg-II line widths of X-ray sources in the E-CDF-S region. BH masses were obtained from scaling relations of \bhm\ with FWHM and luminosity of the broad-line components in the spectra. We refer to the original paper for details on the method and the assessment of the reliability of the masses. The median uncertainty is claimed to be $\sim 0.1$ dex with an additional 0.3 dex due to calibration uncertainties in the scaling relations. This data set was integrated with H$\alpha$ BH mass measurements from \citet{Schulze18}, after re-calibrating the scaling relation to the same one adopted in \citet[][their Eq. (1)]{Suh2015}, and with Mg-II BH mass measurements from \citet{Schramm13}. In total, we collected masses for 40 sources: 35 from \citet{Suh2015}, 3 from \citet{Schulze18} and 2 from \citet{Schramm13}. In the case of multiple BH mass measurements for the same source, we used the average value; however our results do not change if we adopt individual BH mass measurements instead (giving priority to H$\alpha$ measurements). For the subsequent analysis we defined a "robust sample" of 15 sources with available \bhm\ measurements, average $S/N>0.8$ per point and more than 90 points in their X-ray lightcurves in order to have reliable \snxs\ measurements. In fact, we verified that including more sources with lower quality lightcurves does not yield significant improvements in the final analysis and increases the uncertainties on the measured variability (see P17 for further details). 

\begin{figure}
   \centering
   \includegraphics[bb=15 55 715 595, width=0.48\textwidth]{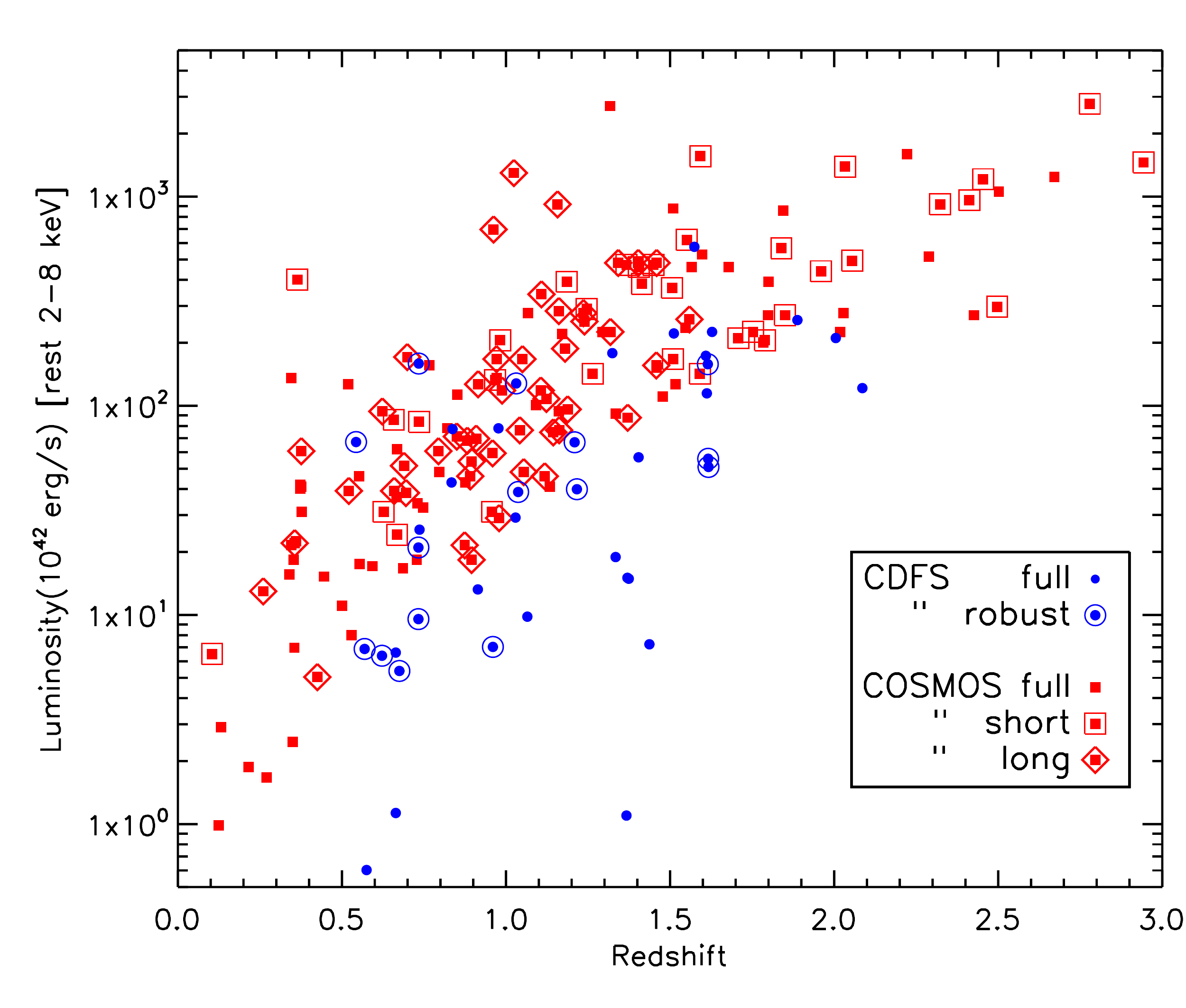}
   \vspace{0.4cm}
   \caption{Luminosity -- redshift distribution of CDF-S and COSMOS sources. Solid symbols represent sources with available BH mass measurements while empty ones highlight the final samples used in this work (see text for details).}
         \label{Lx_z}
\end{figure}

In Fig. \ref{Lx_z} we plot the luminosity--redshift distribution of CDF-S sources with available \bhm\ measurements and those in the robust samples. The sources in the robust sample are distributed between $\sim$ 0.5--1.7 in redshift, $\sim 5\times 10^{42}$--$2\times 10^{44}$ erg s$^{-1}$ in (2--8 keV rest-frame) X-ray luminosity, $10^7$--$10^9$ M$_\odot$ in BH mass, and have low absorption $N_H<2\times 10^{22}$ cm$^{-2}$ as expected from their type 1 spectra.

\subsection{The CDF-S \snxs\ -- \bhm\ relation}
\label{sec:CDFSnxv}

We created \snxs\ -- \bhm\ plots for the CDF-S data using the excess variance measurements of P17 and the BH mass measurements for the sources in the ``robust sample''  discussed in the previous section. P17 calculated the excess variance of the CDF-S AGN using $T_{\rm obs}=45, 128, 654$ and $6005$ day-long (observer's frame) lightcurves (see the discussion in their \S 4.2). Here we used only their excess variance measurements from the 128 and 654 day-long lightcurve segments, i.e. the segments plotted in the two rightmost panels in Fig. 1 of P17. They are separated by almost 5 years, and the excess variance measurements based on them should be uncorrelated. 
On the contrary, the \snxs\ measurements from the longest and shortest timescales use lightcurve parts which overlap with each other (see
Fig. 1 in P17), hence the resulting \snxs\ will be correlated. Note that since the AGN PSD is known to depend on the energy \citep[e.g.][]{McHardy04}, as well as to minimize the effect of absorption, the lightcurves are extracted in the 2--8 keV rest--frame band.

\begin{figure}
   \centering
   \includegraphics[bb=20 25 610 610, width=0.5\textwidth]{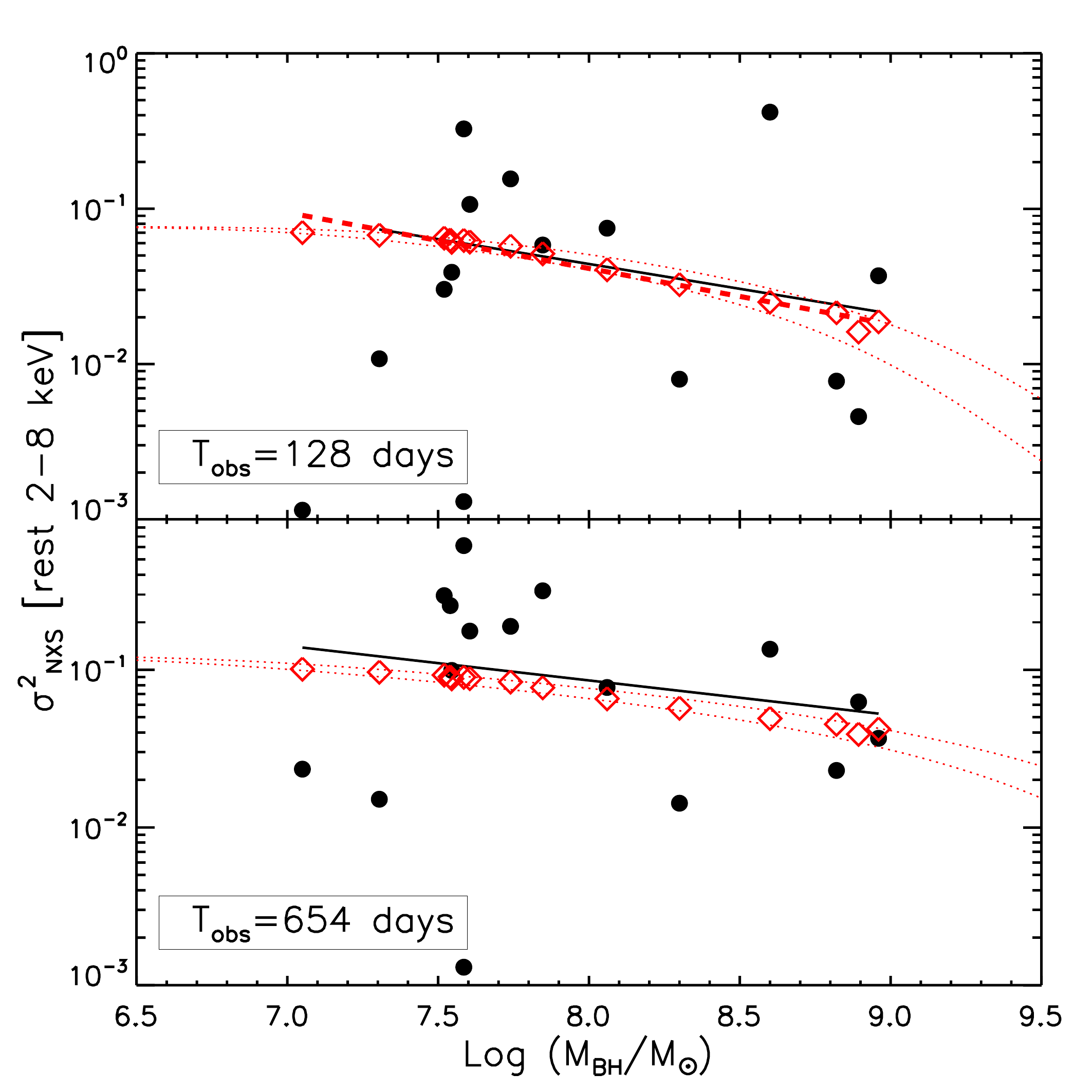}
   \vspace{0.2cm}
   \caption{Plots of \snxs\ vs. \bhm\ for the CDF-S ``robust'' sample (filled circles) for the two different timescales discussed in the text. The solid lines show the linear fit to the data. Open diamonds show the best-fit model predicted variance, computed by taking into account the rest-frame $T_{\rm rest}$, and $\Delta t_{\rm min, rest}$ of each source. The dashed lines mark the best fit to the model predictions (see \S \ref{sec:fitting} for details). The dotted curves represent the model trends for the sources at $z=0.5$ (top one) and $z=2.0$ (bottom one), i.e the approximate redshift range of the CDF-S sample (see Fig. \ref{Lx_z}).}
         \label{fig:varvsbhmcdfs}
\end{figure}

The panels in Fig.\,\ref{fig:varvsbhmcdfs}  show the \snxs\ -- \bhm\ plots for the two different timescales. A weak anti-correlation between \snxs\ and \bhm\ can be observed, although the scatter is considerable. This is due to several reasons. First, the error on \bhm\ and, more importantly, on the individual \snxs\ measurements \citep{Allevato13}, will introduce considerable scatter around the intrinsic relation. In addition, although $T_{\rm obs}$ is the same for all objects, the lightcurve variance depends on the duration in rest-frame, $T_{\rm rest}$, which is not the same for all objects, as their redshifts differ. 

We note that, as we explained in \S \ref{sec:vfptheory}, if both the duration and bin size of the lightcurve are much longer than the bend timescale, then the intrinsic \snxs\ -- \bhm\ relation will be flat (unless the PSD amplitude depends on \bhm). In other words, a lack of correlation between excess variance and BH mass is representative of an intrinsically flat \snxs\ -- \bhm\ relation, thus holding important information regarding the shape of the VFP and the PSD at low frequencies.
 
We cannot use $\chi^2$ to fit the data plotted in Fig.\,\ref{fig:varvsbhmcdfs} (as well as in all figures which show \snxs\ -- \bhm\ relations) and then test whether the model fits the data well, or not, because the excess variance measurements are not Gaussian (and in any case their error is unknown). For that reason, we assumed that a straight line provides a good fit to the data, and we used the ordinary least-squares regression of Y on X (OLS(Y|X)) method of \cite{Isobe90} to fit the data in  Fig.\,\ref{fig:varvsbhmcdfs} (in log-log space), with the linear model defined by Eq.\,(\ref{eq:linemod}), with $\rm \bar{M}=10^8~M_{\odot}$,  which is similar to the average \bhm\ in the CDF-S sample. As such, the line normalization will be best determined at this mass. 
In this case, $\alpha(T_{\rm obs})$ corresponds to the intrinsic variance of an AGN with \bhm\ = $10^8$  M$_{\odot}$, when measured from a lightcurve of duration $T_{\rm obs}$. 

Best-fit results  are listed in Table \ref{table:dataandfit} and black solid lines in Fig.\,\ref{fig:varvsbhmcdfs} show the best-fit linear models. Not surprisingly, given the large scatter of the points around the best-fit lines, the error on the slope is large; however, the error on the normalization is small. This is due to the fact that, given the rather flat best-fit slope, $\alpha(T_{\rm obs})$ is representative of the mean  excess variance of all the points in each plot, which is reasonably well determined from the 15 measurements at each timescale. 

\begin{table*}
\centering
\begin{tabular}{lccccc}
\hline
Survey & $T_{\rm obs}$ & $\Delta t_{\rm min,obs}$ & $\tilde{T}_{\rm rest}[{\rm range}]$ & $\alpha(T_{\rm obs})$ & $\beta(T_{\rm obs})$ \\
            & (days) & (days) & (days)& \\
\hline
 CDF-S & 654 & 0.25 & $334 [\pm 87]$ & -1.07$\pm$0.12 & -0.2$\pm$0.2\\
           & 128 & 0.95 & $65 [\pm 17]$ & -1.36$\pm$0.16 & -0.3$\pm$0.3\vspace{0.2cm}\\
COSMOS &  555 & 0.40 & 240[$^{+88}_{-107}$] & -1.36$\pm$0.10 & -0.16$\pm$0.14 \vspace{0.1cm}\\
~~~~~~"     & 891 & 0.38 & 413[$^{+139}_{-69}$] & -1.29$\pm$0.07 & -0.21$\pm$0.13\vspace{0.2cm}\\
CAIXA & 0.926 & 0.003  & 0.926 & -2.9$\pm$0.2 & -0.71$\pm$0.16\\
... +TARTARUS & 0.463 & " & 0.463 & -2.98$\pm$0.14 & -0.75$\pm$0.14\vspace{0.2cm}\\
long-term \RXTE       & 5110 & 300 & 5110 & -1.38$\pm$0.09 & -0.15$\pm$0.12\vspace{0.2cm}\\
\swift+\RXTE & 9.45 & $\sim$0.5 & 9.45 & -1.81$\pm$0.08 & -0.42$\pm$0.07\\
\hline
\end{tabular}
\vspace{0.4cm}
\caption{Results of the linear fits to the \snxs\ -- \bhm\ relation (in log--log space), for the different datasets used in this work. $T_{obs}$ and $\Delta t_{min,obs}$ represent the maximum and minimum sampled timescales in the \textit{observer} reference frame, while $\tilde{T}_{\rm rest}$ is the median value of the maximum \textit{rest-frame} timescale over all sources; in square brackets we quote the ${T}_{\rm rest}$ range for high-$z$ samples (differences in ${T}_{\rm rest}$ for the low-$z$ samples are not significant). $\alpha(T_{\rm obs})$ and $\beta(T_{\rm obs})$ represent the best-fit intercept and slope of Eq. (\ref{eq:linemod}).}
\label{table:dataandfit}
\end{table*}


\subsection{The COSMOS \snxs\ -- \bhm\ relation}
\begin{figure}
   \centering
   \includegraphics[bb=20 25 610 610,width=0.5\textwidth]{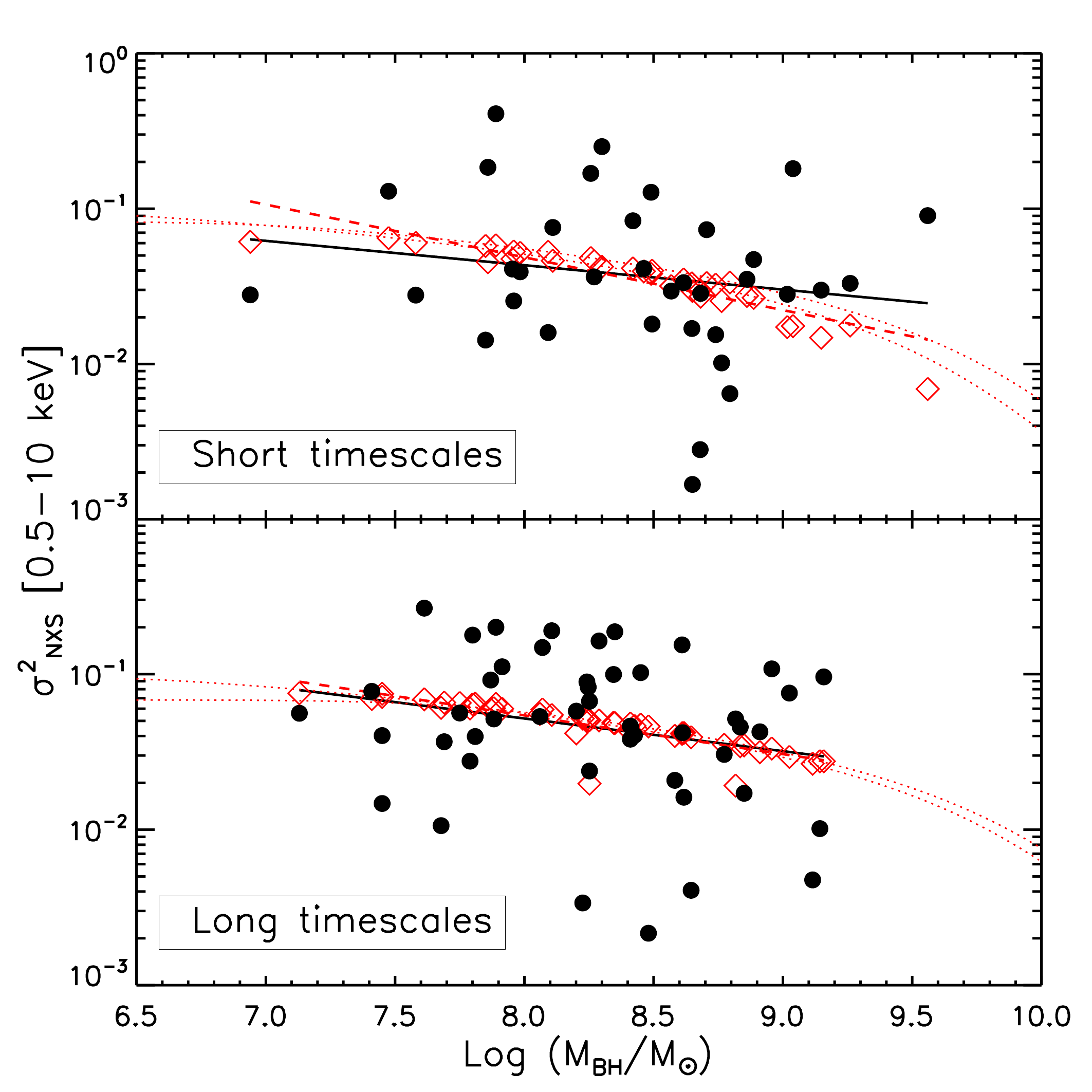}
   \vspace{0.2cm}
   \caption{Plots of \snxs\ vs. \bhm\ for COSMOS sources from \citet{Lanzuisi14}. Symbols have the same meaning as in Fig. \ref{fig:varvsbhmcdfs}.}
   \label{fig:varvsbhmCosmos}
\end{figure}

\citet{Lanzuisi14} used XMM observations in the COSMOS field over a period of $\sim 3.5$ years to study the long-term X--ray variability of a large sample of AGN. We used the data plotted in their Fig. 5 to fit the \snxs\ -- \bhm\ relation. To improve the accuracy of the measured variances, we selected only objects with at least three points in their lightcurves, and with a total rest-frame duration between 100 and 560 days (see their Fig. 1). Since the final sample spans a wide range in $T_{\rm obs}$, we further divided it in two bins based on the rest-frame lightcurve length:  $100~\mbox{days}\leq T_{\rm rest}<330~\mbox{days}$ and $330~\mbox{days}\leq T_{\rm rest}<560~\mbox{days}$, with a median duration $\tilde{T}_{\rm rest}$ of 240 and 413 days respectively. There are 82 AGN in both groups, with a median redshift of 1.5 and 1.0, in the first and second group, respectively (see Fig. \ref{Lx_z}). Their X--ray luminosity ranges from $6\times 10^{42}$ to $3\times 10^{45}$ erg s$^{-1}$ in the 2-10 keV band. We fitted both \snxs\ -- \bhm\ plots with the model defined by Eq. (\ref{eq:linemod}), and the same OLS(Y|X) routine as above (Fig. \ref{fig:varvsbhmCosmos}). The timescales and best-fit results for the COSMOS data are listed in Table \ref{table:dataandfit}.

\section{The variance--BH mass relation of low--redshift AGN}
\label{sec:lowzAGN}
\subsection{Compilation of \snxs\ -- \bhm\ relations from literature}
\label{sec:literaturenxv}

\begin{figure}
   \centering
   \includegraphics[bb=20 25 610 610,width=0.5\textwidth]{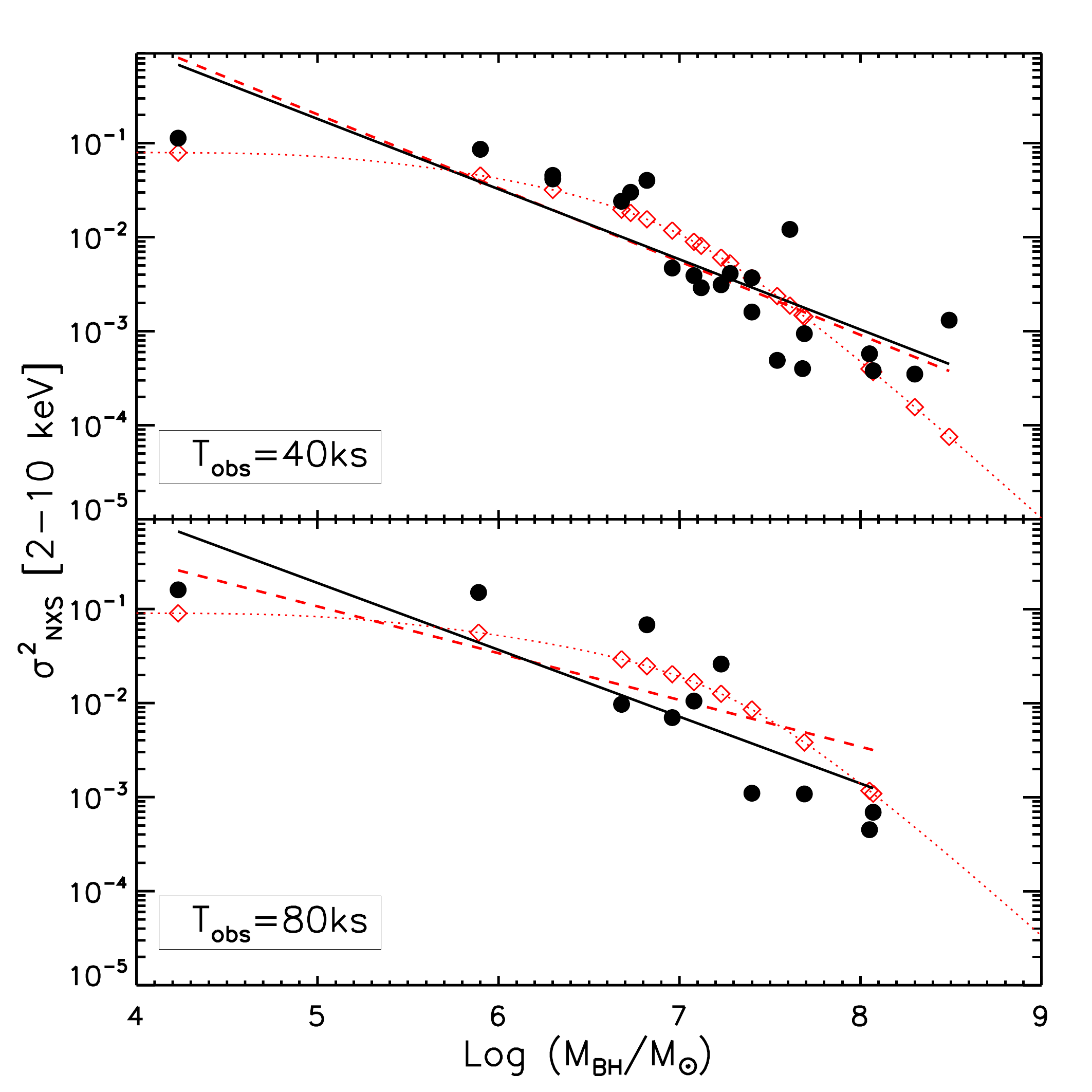}
   \vspace{0.1cm}
   \caption{Plots of \snxs\ vs. \bhm\ for CAIXA+TARTARUS sources with $T_{\rm obs}=40$ks (top panel), and for CAIXA sources with $T_{\rm obs}=80$ks (bottom panel). Data are from \citet{Ponti12} and \citet{Oneill05}. Symbols have the same meaning as in Fig. \ref{fig:varvsbhmcdfs}.}
   \label{fig:varvsbhmCAIXA}
\end{figure}

In order to sample shorter and longer timescales, which are not accessible for high-redshift AGN, we used both published and archival data. 
We first considered the \snxs\ -- \bhm\  data from the CAIXA sample \citep{Ponti12}. These authors presented the results from a systematic study of the excess variance of a large AGN sample using XMM-{\it Newton} lightcurves. We used their \snxs\ measurements (2--10 keV) from the lightcurves with $T_{\rm obs}=80$ ks. They measured the excess variance on three shorter time timescales as well, but the use of the same lightcurves when measuring \snxs\ on different timescales would imply that their \snxs\ measurements would be heavily correlated (for the same source).
We constructed the respective \snxs\ -- \bhm\ plot (see the bottom panel of Fig.\,\ref{fig:varvsbhmCAIXA}) using the data from their ``Rev'' AGN sample.\footnote{The sample with masses derived from reverberation mapping measurements.} In doing so, we updated their BH mass estimates with the measurements listed in the AGN BH mass database \citep{Bentz15}. There are 11 radio-quiet AGN in this sample with excess variance measurements based on 80ks--long lightcurves. For consistency, we fitted the CAIXA \snxs\ -- \bhm\ plot using Eq.\, (\ref{eq:linemod}), with $\rm \bar{M}=10^8\mbox{ }M_{\odot}$, and the same OLS(Y|X) routine that we used to fit the respective CDF-S plots. The best-fit results are listed in Table \ref{table:dataandfit}, and the black solid line in the lower panel of Fig.\,\ref{fig:varvsbhmCAIXA} shows the best-fit line. 


In order to get information on shorter timescales, we considered the excess variance measurements from the TARTARUS sample of \citet{Oneill05}. They used \ASCA 40 ks long lightcurves, and measured the excess variance of nearby, X-ray bright Seyferts (in the 2-10 keV band). We chose 16 (radio-quiet) AGN from their sample with BH mass measurements based on the reverberation mapping technique, as listed in the database of \cite{Bentz15}. We added six sources from the CAIXA sample for which \cite{Ponti12} provide 40 ks \snxs\ measurements (these are: NGC~4151, Mrk~110, Mrk~279, Mrk~590, NGC~4593 and PG~1211+143); they are not part of the 80 ks sample and have BH mass estimates based on reverberation mapping. The respective \snxs\ -- \bhm\ plot is shown in the top panel of Fig. \ref{fig:varvsbhmCAIXA}. As above, we fitted the data using Eq.\, (\ref{eq:linemod}) and the same OLS(Y|X) routine that we used to fit the respective CDF-S plots. Best-fit results are listed in Table \ref{table:dataandfit}, and the solid line in the upper panel of Fig.~\ref{fig:varvsbhmCAIXA} indicates the best-fit line.

\begin{figure}
   \centering
   \includegraphics[bb=20 25 620 323,width=0.5\textwidth]{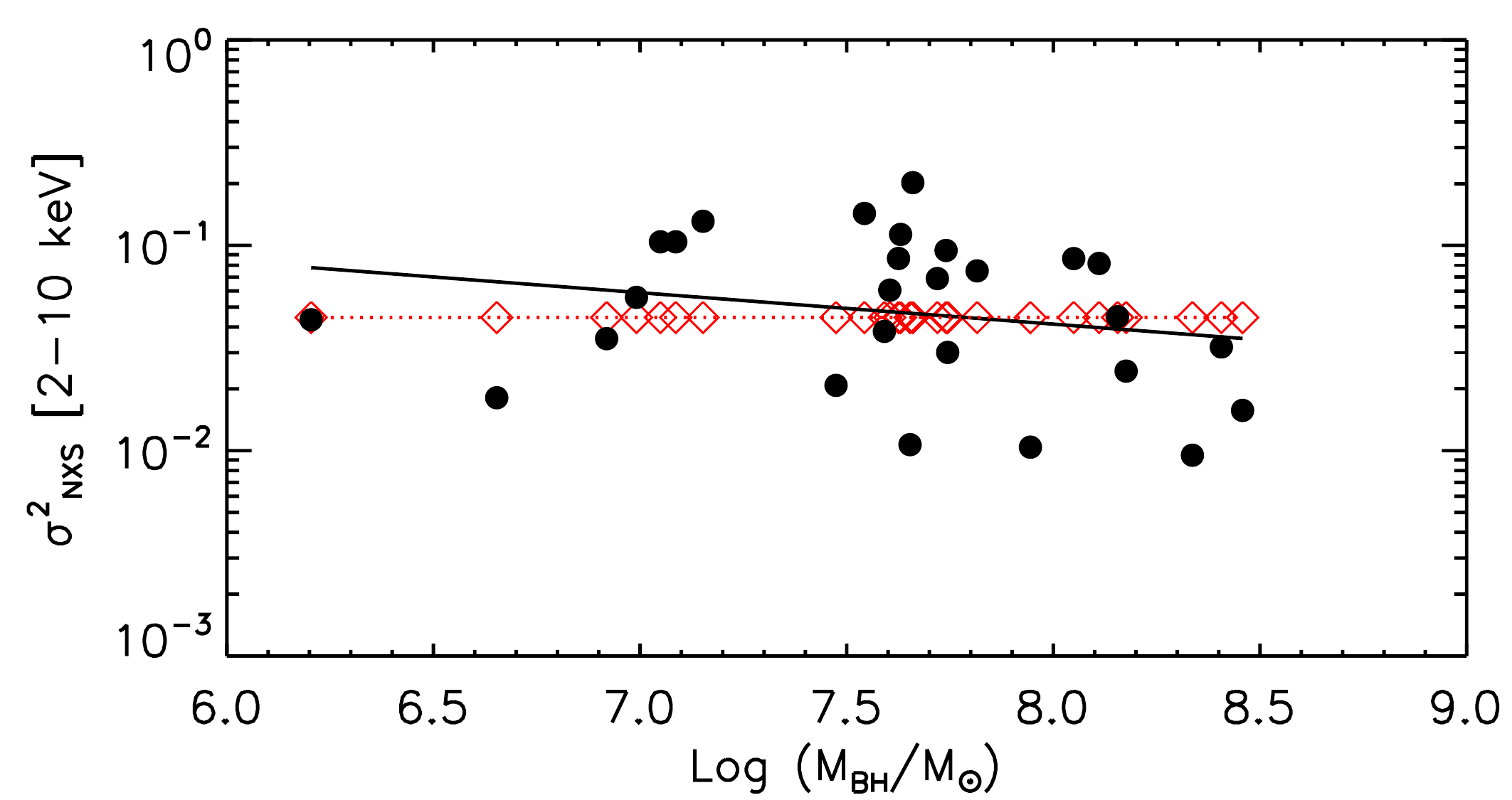}
   \vspace{0.2cm}
   \caption{Plots of \snxs\ vs. \bhm\ for \RXTE\ sources (filled circles) from \citet{Zhang11}. Symbols have the same meaning as in Fig. \ref{fig:varvsbhmcdfs}.}
   \label{fig:varvsbhmRXTE}
\end{figure}

We also considered the \snxs\ -- \bhm\ data from the \RXTE\ lightcurves of \citet{Zhang11}, to get information on longer timescales. They used data from the ASM on board \RXTE\ to study the X-ray variability amplitude of 27 AGN over a period of $T_{\rm obs}=14$ years. Using their \snxs\ and \bhm\ measurements, we fitted the resulting variability--\bhm\ relation with the same model and fitting routine as above. The \snxs\ -- \bhm\ relation, together with the best-fit line, are shown in Fig.\, \ref{fig:varvsbhmRXTE} while the best-fit results are listed in Table \ref{table:dataandfit}. The AGN in the \citet{Oneill05}, \cite{Ponti12} and the \cite{Zhang11} samples are all nearby, and their X--ray luminosity span the range $10^{40}-10^{46}$ erg s$^{-1}$.

\begin{figure}
   \centering
   \includegraphics[bb=20 25 610 320,width=0.5\textwidth]{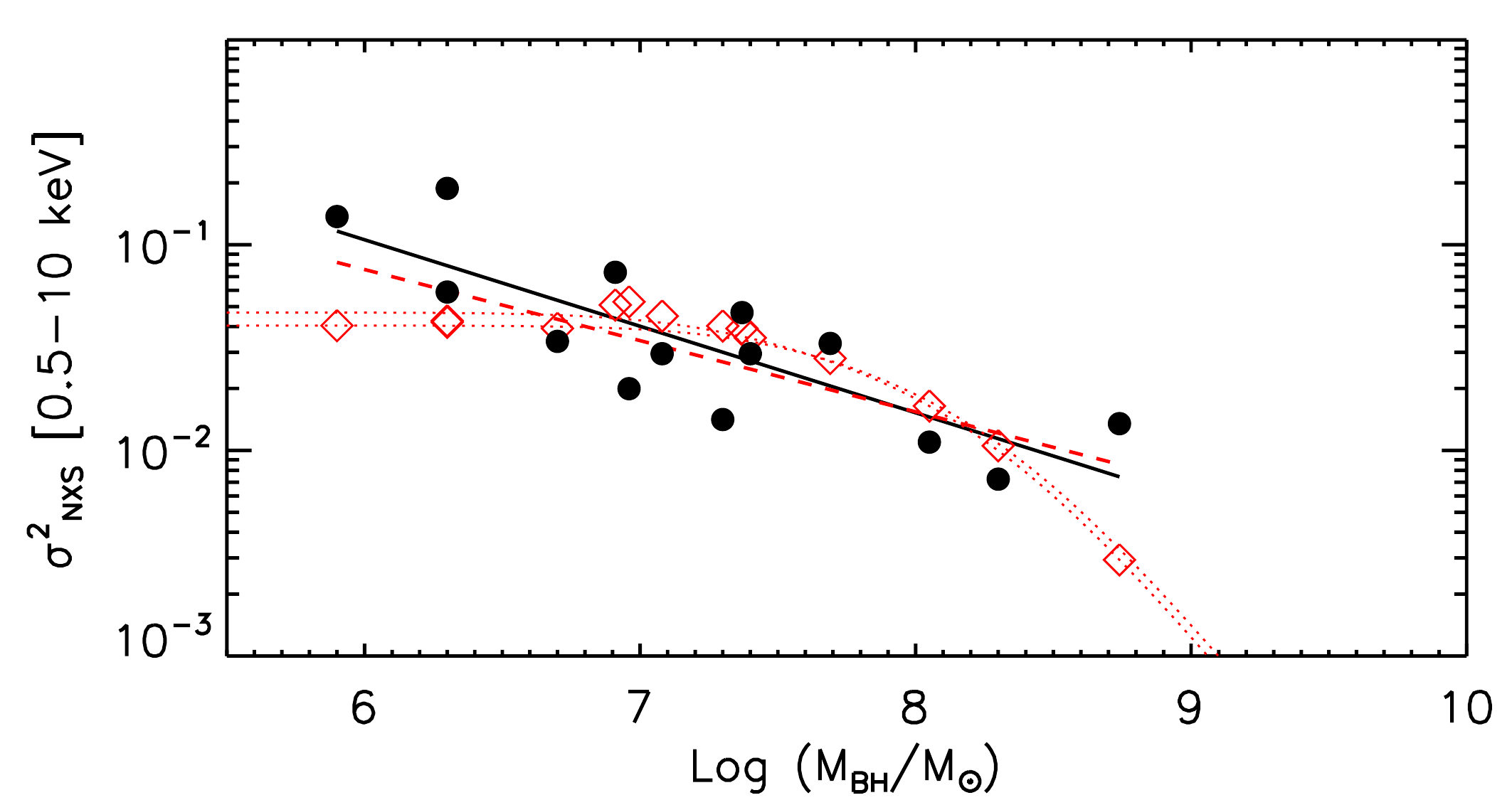}
   \vspace{0.2cm}
   \caption{Plots of \snxs\ vs. \bhm\ for \swift+\RXTE\ sources in Table \ref{table:RXTEandswiftdata}. Symbols have the same meaning as in Fig. \ref{fig:varvsbhmcdfs}.}
   \label{fig:varvsbhmSwiftRXTE}
\end{figure}

\subsection{Medium-term \snxs\ measurements from archival observations}

There is a considerable gap between the duration of the shortest CDF-S lightcurves (128 days) and the $\sim 1$ day-long CAIXA lightcurves. In order to measure the excess variance on intermediate timescales, we used \RXTE/PCA and \swift/XRT data of nearby AGN, and we computed their excess variance using lightcurves which are 9.45 days long (rest-frame). This timescale is ten times longer than the CAIXA lightcurves and $\sim 10$ times shorter than the CDF-S. The sources, together with information on the lightcurves we used and the resulting excess variance, are listed in Table \ref{table:RXTEandswiftdata}.

Column (2) lists the start/end time of the lightcurves we used (in MJD) and, in parentheses, the (minimum) lightcurve bin size $\Delta t_{\rm min}$, in days.  We did not reduce the \swift\ data ourselves; instead we used the lightcurves from \citet{Edelson2019}, for all sources, and \citet{Cackett2020} for Mrk~142.  The \swift\ lightcurves are in the 1.5-10 keV band, except for Mrk~142, where the published lightcurve is in the 0.3--10 keV band. All the other lightcurves were taken from the \RXTE\ AGN Timing \& Spectral Database\footnote{\url{https://cass.ucsd.edu/~rxteagn/}} \citep{Rivers2013}  and they are in the 2-10 keV band. BH mass estimates (from \citealp{Bentz15}) are listed in column (3).

We divided the lightcurves into segments with a (rest-frame) duration of 9.45 days (as determined by the shortest lightcurve in the sample). We computed the excess variance of each segment in the usual way (i.e., using Eq. (1) in P17). We then computed the mean of the individual \snxs\ measurements for each source, which is listed in the last column of Table \ref{table:RXTEandswiftdata}. As for all the other samples, we fitted the \snxs\ -- \bhm\ relation using Eq.\,(\ref{eq:linemod}) and the OLS(Y|X) routine.
The result is shown in Fig. \ref{fig:varvsbhmSwiftRXTE}, together with the best-fit model, while the best-fit results are listed in Table \ref{table:dataandfit}.


\begin{table}
\caption{The \RXTE\ and \swift\ lightcurves we used to measure the excess variance of AGN on timescales of $\sim 10$ days.}
\label{table:RXTEandswiftdata}
\begin{tabular}{llcc} 
\hline
Name$^*$ &          Dates($\Delta t_{\rm min,obs}$, in days)  & log(M$_{\rm BH}$) & log(\snxs) \\   
     &          (MJD)             &   (M$_{\odot}$) &                       \\
\hline
F9	            &$52145.0  - 52179.0~(0.14)$       &8.3  & -2.14	\\	   
PG~0804	        &$53300.3  - 53362.8~(0.48)$       &8.74 & -1.87	\\
Mrk~110	        &$53695.5  - 53760.5~(0.27)$       &7.30 & -1.85	\\
NGC~3227	    &$51258.6  - 51300.4~(0.76)$       &6.7  & -1.47	\\
Mrk~142	        &$58484.4  - 58603.9~(0.65)$       &6.3  & -0.73	\\
NGC~3516	    &$50523.0  - 50657.9~(0.56)$       &7.40 & -1.53	\\
NGC~3783	    &$51960.1  - 51980.1~(0.13)$       &7.08 &		\\
    	        &$54504.2  - 54617.1~(0.34)$       &     & -1.53	\\
NGC4051	        &$51627   - 51637.5~(0.44)$        &5.9  &        \\
                &$51665.4  - 51730.1~(0.26)$       &     &        \\
	            &$54147.5  - 54287.2~(1.03)$	      &     & -0.86	\\
NGC4151(S)	    &$57438.0 - 57505.8~(0.2)$         &7.37 & -1.33	\\
NGC4593(S)	    &$57582.8  - 57605.4~(0.12)$       &6.91 &         \\
	            &$53701.4  - 53766.7~(0.27)$       &     & -1.13	\\
MCG-6-30-15	&$50318.3  - 50355.6~(0.62)$           &6.3  &        \\
                &$51378.1  - 51388.2~(0.1)$        &     &        \\
	            &$51622.7  - 51688.6~(0.25)$	      &     &        \\
	            &$54261.1  - 54329.1~(1.06)$	      &     & -1.23	\\
NGC5548(S)      &$56706 - 56833.6~(0.48)$          &7.69 & -1.48	\\
Mrk~509(S)	    &$57829.9  - 58102.5~(1.07)$       &8.05  & -1.96	\\
NGC7469	        &$50244.1 - 50276~(0.1)$           &6.96  & -1.7	\\
\hline
\end{tabular}
* The letter (S) after a source name indicates the use of \swift\ lightcurves.
\end{table}

\section{The observed VFP of AGN}
\label{sec:observedVFP}

\begin{figure}
   \centering
   \includegraphics[bb=0 25 605 595, width=0.5\textwidth]{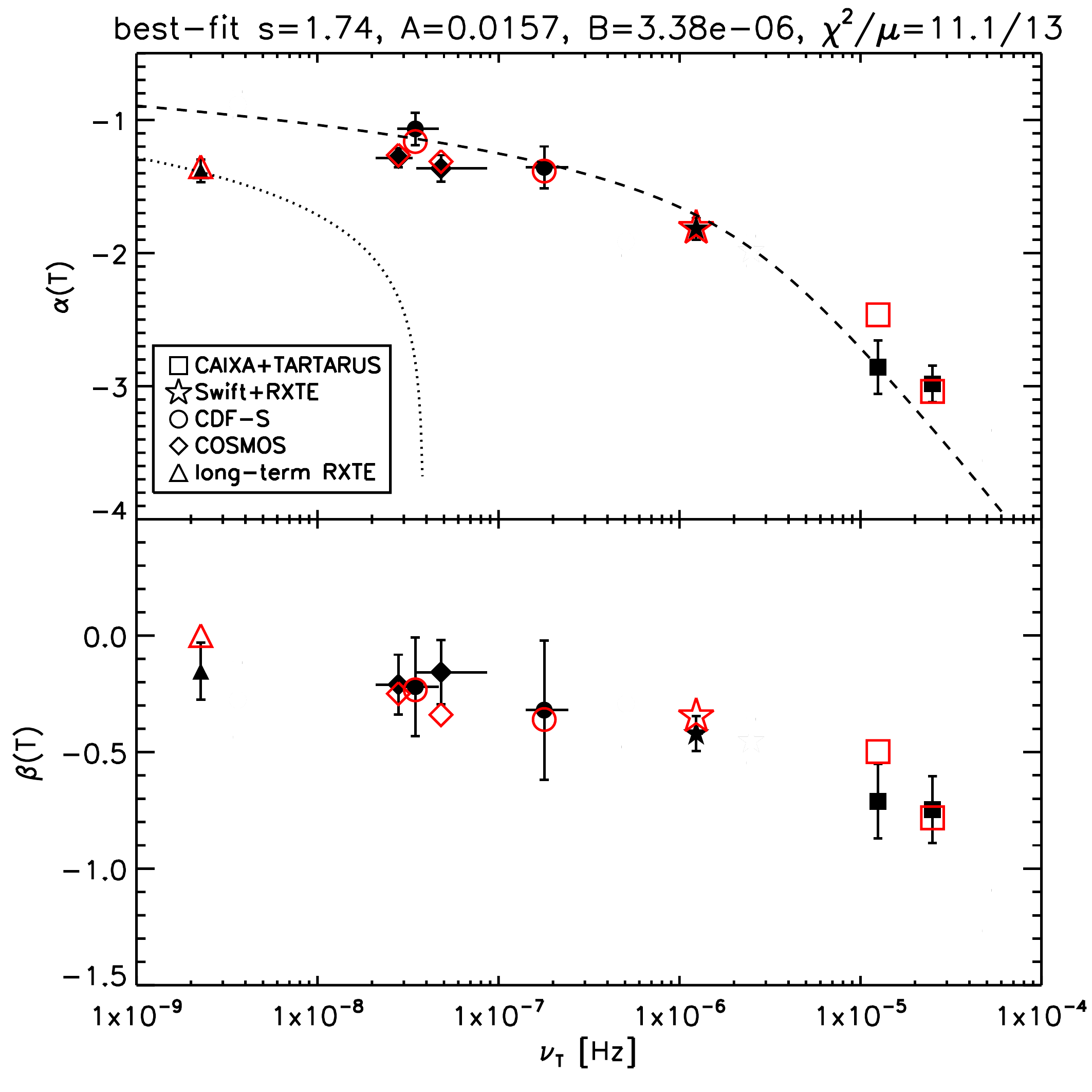}
   \vspace{0.1cm}
      \caption{{\it Upper panel:} 
The observed VFP of a $10^8$ M$_{\odot}$ AGN as a function of the rest-frame frequency $\nu_T$, using our measurements, from Table \ref{table:dataandfit} (filled symbols). The red empty symbols indicate the best-fit model values ($\alpha(T_{\rm max})$ in Eq. (\ref{eq:linemod})). For reference we also show the theoretical expected trends for a $10^8$ M$_\odot$ BH monitored with an average $\Delta t_{\rm min}=250$ s (dashed line) and $\Delta t_{\rm min}=300$ days (dotted line), but we stress that the comparison should be done between the filled and empty points. {\it Lower Panel:} Measured and best-fit model slope of the \snxs--\bhm\ relations ($\beta(T_{\rm max})$ in Eq. (\ref{eq:linemod})); symbols have the same meaning as those in the upper panel.} 
         \label{fig:superpsd}
\end{figure}

As we already explained, the best-fit $\alpha(T)$ values listed in Table \ref{table:dataandfit} are representative of the variance of a 10$^8$ M$_\odot$ AGN, on the timescales that are listed in the second column of the same table. 
We therefore used the $\alpha(T)$ values listed in this table and we constructed the VFP for the 10$^8$ M$_{\odot}$ AGN.
Figure \ref{fig:superpsd} shows the best-fit $\alpha(T)$ and $\beta(T)$ values plotted 
as a function of \nT\ (top and bottom panels, respectively). The two timescales probed by the CAIXA+TARTARUS data (filled squares) constrain the high frequency part of the VFP, and the \swift+\RXTE\ data (filled stars) allow us to sample intermediate timescales. The results from CDF-S (filled circles) provide information on long timescales, while the COSMOS and the \RXTE\ points (filled diamonds and filled triangle, respectively) further improve the accuracy of the VFP at low frequencies.

The limits of the PSD integral in Eq.\,(\ref{eq:svsnumod}) depend on rest-frame $T_{\rm max}$ and $\Delta t_{\rm min}$. All AGN in the CAIXA, TARTARUS, \RXTE\ and \swift\ samples are nearby and all lightcurves are of the same duration, hence $T_{\rm obs}=T_{\rm rest}$ (and $\Delta t_{\rm min,obs} = \Delta t_{\rm min,rest}$) in these samples.  
Things are more complicated in the CDF-S and COSMOS samples. $T_{\rm obs}$ is the same for all AGN in the CDF-S; however, $T_{\rm rest}$ is not, because they do not have the same redshift. In the case of the COSMOS AGN, even $T_{\rm obs}$ differ for different sources due to the survey strategy \citep[see][]{Lanzuisi14}.  Due to differences in $T_{\rm rest}$ in these samples, 
in Fig.~\ref{fig:superpsd} we plot all quantities as a function of \nT\ = $1/\tilde{T}_{\rm rest}$, where $\tilde{T}_{\rm rest}$ is the median lightcurve duration in the AGN rest-frame
($\tilde{T}_{\rm rest}$ values are listed in the fourth column of Table \ref{table:dataandfit}). Note that this choice is for visualization purposes only, since in the fitting procedure we properly take into account the actual rest-frame timescales sampled for each individual source (see \S,\ref{sec:fitting}).  

The overall VFP (upper panel in Fig.~\ref{fig:superpsd})  appears to be very well described by a single function, from the lowest to the highest sampled frequencies. 
We can see the logarithmic rise of the  variance with decreasing frequency (i.e., increasing timescale), with a slope which is roughly equal to $-1$, from the CAIXA+TARTARUS to the high-frequency CDF-S point. This implies a PSD slope of $\sim -2$ at high frequencies. The variance--frequency slope flattens at lower frequencies, indicating the presence of a bend frequency, analogous to the PSD bend frequency, $\nu_b$, somewhere between (1--20 days)$^{-1}$. The low and high-frequency parts of the observed VFP are determined by the high$-z$ and low$-z$ AGN samples, respectively, but the important observational result is that the low-frequency VFP appears to be the continuation of the high-frequency VFP, without any hints of a normalization mismatch between the two parts. 

At low frequencies, the \RXTE\ variance may appear to underestimate the value expected from a simple extrapolation of the CDF-S and the COSMOS measurements in the VFP,
although $T_{\rm rest}$ of the \RXTE\ lightcurves is longer than the rest-frame duration of the longest CDF-S lightcurves. In principle, this could suggest that the PSD normalization at low frequencies is not the same in the distant and local AGN. However, $\Delta t_{\rm min}$ in the \RXTE\ lightcurves is also significantly larger than the (rest-frame) minimum timescale in all other lightcurves. In fact, it is almost certainly much longer than the PSD bend frequency of a 10$^8$ M$_{\odot}$ AGN. This implies that we are missing a significant part of the intrinsic variance in the \RXTE\ lightcurves, hence the smaller $\alpha(T)$ value. 

The bottom panel in Fig.~\ref{fig:superpsd} shows how the slope of the \snxs--\bhm\ relation varies with frequency. As with the VFP plot, the best-fit slopes at high frequencies appear to connect smoothly, without any normalization discontinuities, with the best-fit $\beta(T)$ values at lower frequencies. The slope of the \snxs--\bhm\ plots at low frequencies (long timescales) approaches zero, i.e. the bending frequency of the 10$^8$ M$_{\odot}$ AGN is probably higher than both the mean bin size and duration of the available lightcurves in the \RXTE\ sample and, to some extend, in the CDF-S and the COSMOS sample. 

\section{Model fitting procedure and best-fit results}
\label{sec:fitting}



We considered a grid of $A$, $B$, and $s$ values, and for each $(A,B,s)$ combination we used Eqs.\,(\ref{eq:nubmod}) and (\ref{eq:svsnumod}) to compute the variance $\sigma^2$ for the AGN in the CDF-S, COSMOS, CAIXA+TARTARUS, \RXTE\ and \swift+\RXTE\ samples. To this end we derived $\nu_T$, $\nu_{max}$ and $\nu_b$ using the \bhm\ and $z$ values of each source, and the timescales $T_{\rm obs}$ and $\Delta t_{\rm min,obs}$ in Table \ref{table:dataandfit} (we assumed $z=0$ for the AGN in the CAIXA+TARTARUS, \RXTE\ and the \swift+\RXTE\ samples which only contain nearby AGN).

In order to properly fit the data we must take into account the biases due to a sparse and/or irregular sampling of the lightcurves. To this end we used Eq. (11) in \citet{Allevato13} to compute the bias affecting the measured \snxs\ and we include the same factor in the model variance $\sigma^2$. The bias correction depends on the PSD slope itself, so we used the ``average'' slope (from Eq.(\ref{eq:PSDmod})) over the range of sampled rest-frame timescales for each source.
Furthermore we assumed a sparse sampling for the COSMOS sources, 
and a continuous sampling pattern for the remaining samples.

In this way, for each model parameter combination we ended up with a pair of model $(\sigma^2,$ \bhm) values, for each source and timescale in the 
 CDF-S, COSMOS, the CAIXA+TARTARUS, and the \swift+\RXTE\ samples. 
If the observed \snxs\ measurements were Gaussian variables with known errors, then we could use $\chi^2$ statistics to fit all the observed \snxs\ versus \bhm\ data plotted in Figs.~\ref{fig:varvsbhmcdfs} -- \ref{fig:varvsbhmSwiftRXTE} with the model $(\sigma^2,$ \bhm) curves. However this is not the case, so we cannot fit directly the model to the observed excess variance -- BH mass plots. Instead, we followed a different way to fit the model to the data.


For each model parameter combination we fit the resulting $\sigma^2$--\bhm\ points with Eq.\,(\ref{eq:linemod}), with $\rm \bar{M}=10^8\mbox{ }M_{\odot}$, using the same OLS(Y|X) routine that we used to fit the data. In the end, each set of model parameters, $(A,B,s)$, would result in two best-fit values, $\alpha_{mod}(A,B,s,T_{\rm obs})$ and $\beta_{mod}(A,B,s, T_{\rm obs})$, for each  $\sigma^2$--\bhm\ relation.

\underline{{\it The overall model fit:}} Based on what we discussed above, each set of model parameters, $(A,B,s)$, would result in eight $\alpha_{mod}(A,s,B, T_{\rm obs})$ and $\beta_{mod}(A,s,B, T_{\rm obs})$ values, i.e. one for each of the 
observed \snxs\ --\bhm\ relations plotted in Figs.\,\ref{fig:varvsbhmcdfs} -- \ref{fig:varvsbhmSwiftRXTE}.
To get the best-fit model, we minimized  $\chi^2$, defined as follows: 
\begin{equation}
\begin{split}
\label{chi2}
 \chi^2=\sum_{i=1}^N \ \Biggl\{\left[\frac{\alpha_{mod}(A,B,s, T_{\rm obs, i}) - \alpha(T_{\rm obs,i})}{\delta [\alpha(T_{\rm obs,i})]}\right]^2 \\
 + \left[\frac{\beta_{mod}(A,B,s, T_{\rm obs, i}) - \beta(T_{\rm obs,i})}{\delta [\beta(T_{\rm obs,i})]}\right]^2\Biggr\},
\end{split}
\end{equation}

\noindent where $\delta [\alpha(T_{\rm obs,i})]$ and $\delta [\beta(T_{\rm obs,i})]$ are the errors on the best-fit normalization and slope values, $\alpha(T_{\rm obs})$ and $\beta(T_{\rm obs})$, listed in Table \ref{table:dataandfit}. 
Defined in this way, the best-fit model is the one that minimizes the differences between the model and the observed VFP (i.e. the points plotted in the upper panel of Fig.\, \ref{fig:superpsd}), as well as the model and the observed slope of the \snxs--\bhm\ relations (bottom panel of Fig.\, \ref{fig:superpsd}).





\begin{figure}
   \centering
   \includegraphics[bb=0 25 605 595, width=0.5\textwidth]{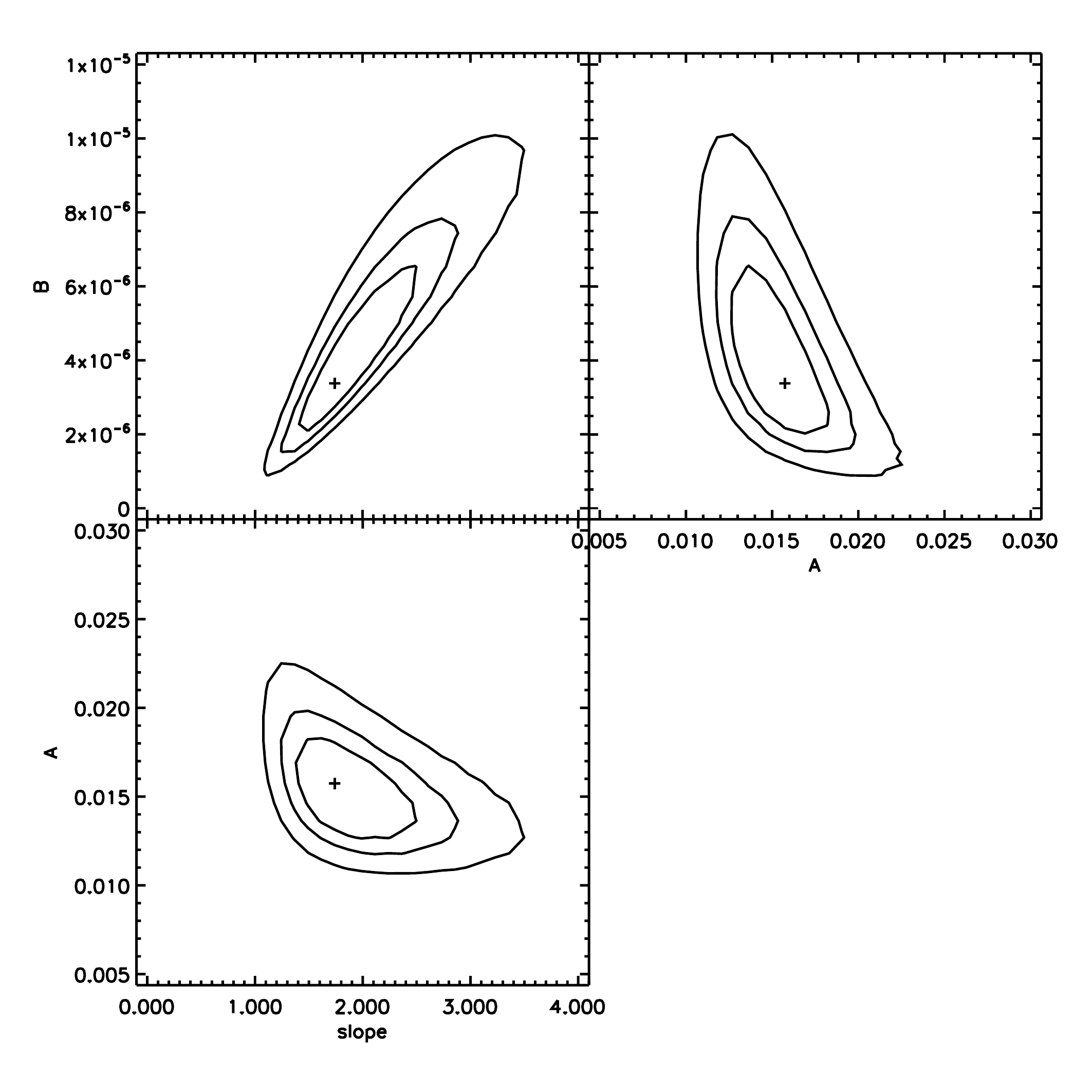}
      \caption{Best-fit values (crosses) and 68\%, 90\%, 99\%  uncertainty contours for each combination of two fitted interesting parameters A, B (i.e. $\nu_b$ for a $10^8\ {\rm M}_{\odot}$ BH) and slope $s$. 
         \label{fig:superpsderr}}
\end{figure}

The best-fit results are: $\chi^2_{\rm min}=11.1$ for $\mu=13$ degrees of freedom, corresponding to $P(<\chi^2,\mu)=0.6$, at $A=0.016^{+0.002}_{-0.003}$ Hz$^{-1}$, $B=3.4_{-1.4}^{+3.1} \times 10^{-6}$ Hz, and $s=1.7_{-0.4}^{+0.9}$. Here the errors are the 90\% uncertainties for two interesting parameters. The errors of the best-fit parameter values are both asymmetric and correlated as shown in Fig. \ref{fig:superpsderr}. Open diamonds in Figs.\,\ref{fig:varvsbhmcdfs} -- \ref{fig:varvsbhmSwiftRXTE} show the best-fit, model ($\sigma^2,$\bhm) predictions for each source in these plots. The dashed red lines show the best linear fit to the model points. Clearly, in some cases (e.g. Figs.~\ref{fig:varvsbhmCAIXA} and \ref{fig:varvsbhmSwiftRXTE}) the model predicts a curved \snxs\ -- \bhm\ relation (dotted curve), which actually may be closer to the observed one. But, even in these cases, a straight line can provide a reasonably good fit to the data (and the model predictions). Since we fit a straight line to both the observed and the model relations, we can compare the model predicted and the observed best-fit line parameters, and search for the parameter values which provide the best agreement between data and the model predictions. The open symbols in Fig.\, \ref{fig:superpsd} show the normalization and slope of the best-fit lines to the model \snxs\ -- \bhm\ relations that are closest to the data. 


Strictly speaking, the VFP plotted in the upper panel of Fig.\,\ref{fig:superpsd} is representative of the average VFP of an AGN with a mass of 10$^8$ M$_{\odot}$. 
The best-fit  parameters also refer to such an object. For example, if we had normalized the \snxs\, vs. \bhm\, relation at a different \bhm, then the expected VFP would be different (see the black and red lines in Fig.~\ref{fig:vfp}) and the best-fit bending frequency would be different if $\nu_b$ depended on BH mass. However, 
the fact that we consider the VFP of an AGN at 10$^8$ M$_{\odot}$ is merely a "technicality" as this mass is close to the mean BH mass of the sources in the samples we considered. In reality, the best-fit $\alpha_{mod}(A,B,s,T_{obs}$) and $\beta_{mod}(A,B,s,T_{obs}$) model parameters are computed by fitting a straight line to all the model points (i.e. open red points in Figs.~\ref{fig:varvsbhmcdfs} - \ref{fig:varvsbhmSwiftRXTE}), Hence, the best-fit model VFP plotted in Fig.\, \ref{fig:superpsd} holds information about the PSD of all AGN in each sample, and the same would be true if we had normalized the best-fit lines to another BH mass.

We find it impressive that a {\it single} PSD model can fit the observed VFP so well, given that we constructed \snxs\ -- \bhm\ plots using lightcurves of 160 AGN, nearby and distant, observed with different satellites, with different sampling and duration. This result strongly suggests that, on average, the X-ray PSD, over five orders of magnitude in frequency (i.e. from timescales of $\sim 40$ ks up to $\sim 10-15$ years) is described by the {\it same} form for all AGN at $z\leq 2-3$. If there were significant differences between the amplitude, and/or bend frequency, or high-frequency slope of the high$-z$ and low$-z$ PSDs, then the low and high frequency parts of the VFP plotted in the top panel of Fig.~\ref{fig:superpsd}, which are determined by the high$-z$ and low$-z$ AGN respectively, would differ significantly (see for example the various curves in Fig.~\ref{fig:vfp}). 

\section{Summary and discussion}
\label{sec:discussion}

We used excess variance measurements computed using short (CAIXA+TARTARUS, \citealp[][]{Ponti12,Oneill05}), intermediate (\swift+\RXTE, this work) and long (\RXTE,  \citealp{Zhang11}) timescale lightcurves, as well as lightcurves from the COSMOS and the CDF-S surveys \citep[P17]{Lanzuisi14}, to construct \snxs\ -- \bhm\ plots on various timescales. We fitted them with a simple linear  model (in the log-log space), and we studied the resulting variance-frequency plot (VFP), together with the slope of the variance--BH mass relations as a function of frequency. Our main result is that the hypothesis of a common X--ray PSD form 
in all AGN, which remains the same irrespective of redshift and luminosity, is fully consistent with the observed VFP. 

The variance vs. timescale relation of the CDF-S sources was used in P17 as well (see their Fig.\, 7), to show that the excess variance measurements were indeed consistent with the assumption of a bending power-law PSD, and in \citet{Zheng2017} to study the low-frequency PSD slope. In this work, we consider a much larger data set to create a more detailed VFP, and we use it together with the slope of the variance -- BH mass plots, to actually constrain the X-ray PSD. 

We use excess variance measurements from over a hundred AGN, with a wide range of BH mass ($\sim 10^6-10^9$ M$_{\odot})$ and X--ray luminosity ($\sim 10^{40}-10^{46}$ erg s$^{-1}$) to determine the VFP over timescales from a few hours up to 5 -- 14 years. Based on the fact that the VFP and the PSD hold the same information, we were able to determine the average PSD of the AGN in our sample. Detailed PSDs have been determined (and fitted with models) only for nearby AGN. To the best of our knowledge, this is the first time that the average X--ray PSD of a representative sample of (X--ray selected) AGN is accurately determined, up to a redshift of $\sim 3$. 

The fact that the observed VFP is well fitted by a single function is remarkable. 
The average variance, $\alpha(T)$,
has been measured using lightcurves of different objects, obtained from different instruments, at different times, and over entirely different timescales. Our results thus strongly suggest that the shape of the average PSD (defined by eq.~\ref{eq:PSDmod}) is the same in all AGN, irrespective of luminosity and/or redshift,
and is consistent with almost all nearby AGN. We note that the PSD shape of least one nearby AGN, namely Ark 564, is different. It has a power law shape and shows two breaks, at high and low frequencies \citep{Papadakis02, McHardy07}. This shape implies that Ark 564 may be in a state which is equivalent to the so-called ``high/intermediate state" in black hole X--ray binaries. Our results suggest that such a state should be rare among AGN.

Regarding the low-frequency slope we assumed a fixed value. If we consider a more general extension of Eq.\,(\ref{eq:PSDmod}), i.e.\, $\mbox{PSD}(\nu)=A\nu^{-l}\left [ 1+\left(\frac{\nu}{\nu_b}\right)^{(s+l)} \right ]^{-1}$, this means fixing $l = -1$. On the other hand \citet{Zheng2017} suggested that a steeper low-frequency slope $l\simeq -1.2$ is more appropriate to fit the CDF-S data. In order to test this possibility we repeated all our fits using the more general expression above; we find that our data are better fit by the canonical slope of $-1$ and, while we cannot rule out a marginally steeper low-frequency slope, a value as steep as -1.2 is excluded at the 99\% significance level.

Our results are not meant to imply that all AGN should have the same high frequency slope and PSD amplitude. Most likely, both the PSD amplitude and high frequency slope will be distributed over a range of values, and the  best-fit values we report should be indicative of the means of these distributions. For example,  \cite{Gonzalez-Martin2012} studied in detail the X--ray PSD in X--ray bright, local AGN and their results do show a rather broad range of high-frequency slopes (between 1.8 and 4.6, see their Table 4). Similarly, their best-fit PSD amplitude values (also listed in their Table 4) show a broad range of values, from 0.001 to 0.04. What is interesting is that the means of these distributions are the same as our results. 


Indeed, the best-fit, high frequency PSD slope from the modeling of the mean VFP is $-(1+s)=-2.7_{-0.4}^{+0.9}$ ($90\%$ confidence limits). Although the PSD slope may depend on the energy band \citep[e.g.][]{McHardy04}, we point out that in our case we used whenever possible the hard X-ray band (i.e., $E\gtrsim 2$ keV) and even in the COSMOS sample the large median redshift of the sources implies that we tend to sample energies $\gtrsim 1$ keV. 
The median of the high frequency PSD slopes reported by \cite{Gonzalez-Martin2012} in their Table 4 is -2.57, for sources where a bend frequency has been detected\footnote{In case of multiple entries, we adopted the best-fit results of \cite{Gonzalez-Martin2012}, except from Ark 564 and PKS0558-504. We have adopted the results of \cite{McHardy07} and \cite{Papadakis10} for these sources, because these authors used more data sets, on various timescales, to compute the PSDs.}. This is  consistent with our results.

In addition, our best-fit PSD normalization of $A=0.016^{+0.002}_{-0.003}$ Hz$^{-1}$, is also consistent with the mean PSD normalization as determined by the PSD modeling of local AGN. According to Eq.\,(\ref{eq:PSDmod}), the PSD amplitude at the bend frequency, in terms of $\nu_b \times \mbox{PSD}(\nu_b)$, is equal to $A/2$. Based on our best-fit results, this is $0.008\pm 0.001$, which is fully consistent with the mean PSD amplitude of $\sim 0.009$ reported by \cite{Gonzalez-Martin2012}\footnote{Note that the amplitude reported in  \cite{Gonzalez-Martin2012} is $0.009\pm0.011$ where the uncertainty represents the standard deviation of their sample; using the proper error on the mean their measurements yield $0.009\pm 0.003$ which is still consistent with our result. Even removing from their sample the 3 NLSy1 with very high normalization, they would obtain $0.0044\pm0.0013$, consistent within $2\sigma$ with our value.}

The broad-band VFP plotted in the upper panel of Fig.\,\ref{fig:superpsd} shows a clear flattening below $\sim 10^{-6}$ Hz, which implies (from our best-fit result) a PSD low frequency bend at $\nu_b=3.4_{-1.4}^{+3.1} \times 10^{-6}$ Hz. This  corresponds to a bend timescale of $(1.8-5.8)$ days (90\% confidence). 
According to \cite{Gonzalez-Martin2012}, log$(T_b)\simeq$ log(\bhm) $-1.7$,
where $T_b$ is in days and \bhm\ is the BH mass in units of 10$^6$ M$_{\odot}$.\footnote{We have assumed that the constant $A$ in Eq.\,(4) of \cite{Gonzalez-Martin2012} is equal to 1, which implies that $T_b$ is proportional to \bhm.}
For a 10$^8$ M$_{\odot}$ AGN, this relation predicts $T_b=2$ days, which is consistent with our results. 
On the other hand, according to \cite{McHardy06}, the bend timescale should also depend on the source luminosity. These authors find that log$(T_b)=2\mbox{ }\times$ log(\bhm) $-$ log(L$_{\rm bol})-2.33$,
where $T_b$ is in days, \bhm\ is the BH mass in units of 10$^6$ M$_{\odot}$, and L$_{\rm bol}$ is the bolometric luminosity in units of 10$^{44}$ erg s$^{-1}$.\footnote{These are the best-fit results for the combined AGN and Cyg X-1 sample.} We have assumed that the constants $A$ and $B$ in the \cite{McHardy06} equation are equal to 2 and 1, respectively, which means that the bend timescale is proportional to the BH mass and inversely proportional to accretion rate (in units of the Eddington limit). For $T_b=3.5$ days, as derived in this work, and \bhm\ = 10$^8$ M$_{\odot}$, then L$_{\rm bol}=1.3\times 10^{45}$ erg s$^{-1}$, which is 10\% of L$_{\rm Edd}$. The VFP analysis thus allows for a dependence of the PSD bend frequency on the AGN accretion rate as well. 

We note that our results suggest that the \textbf{average} bend frequency and PSD amplitude are the same for the high$-z$ and the low$-z$ sources. This implies that either these two PSD parameters do not depend on accretion rate, or that the accretion rate is the same for both the nearby and the distant AGN in our sample. P17 found that the average accretion rate of the CDF-S sources is $\sim 0.05-0.1$ of the Eddington limit. Regarding the low$-z$ objects, if we use the data listed in Table 1 of \citep{Zhang11} for the \RXTE\ sample (which are representative of all our low$-z$ AGN) we find an average accretion rate of $\sim 0.06$ of the Eddington limit, which is  comparable with the accretion rate of the CDF-S sources. We will need to study the VFP of AGN with significantly different accretion rates, in order to investigate the dependence of the PSD amplitude and bending frequency on the accretion rate.

Our results should be useful in future variability studies of large AGN samples, using X--ray lightcurves from, e.g, the eROSITA all-sky survey, and future surveys that may be conducted with the eXTP \citep{Zand2019}, Einstein probe \cite{Yuan2018,Yuan2022} and Star-X \citep{Saha17} proposed missions, as well as Athena \citep{Nandra2013}, Lynx \citep{Gaskin18} or AXIS \citep{Mushotzky18}, provided that deep surveys are properly planned to probe the time domain as well \cite[see e.g.][]{Paolillo12}. Previous studies, such as P17, \textit{assumed} that the shape of the X-ray PSD was the same both in nearby and distant, luminous AGN. We show in this work that this is indeed the case. Thus our result can be useful to any study that involves the modeling of the ensemble X--ray variability of AGN in order to, e.g., use them as cosmological probes \citep[][]{LaFranca2014,Lusso2019,Lusso2020,Demianski2020}, or to constrain the AGN demographics through their ensemble X-ray variability properties \citep{Sartori2019, Georgakakis2021}. From a more physical point of view, a common X--ray PSD shape implies that the same variability mechanism operates in all luminous AGN, and that the mechanism does not evolve with time until up to at least $z\sim 2-3$. This result is also in agreement with the lack of evolution observed in AGN spectral features, such as the UV vs. $L_X$  ratio $\alpha_{ox}$ \citep[e.g.][]{Lusso2010,Lusso2020} or the X-ray spectral slope $\Gamma$ \citep[e.g.][]{Young2009}, and indicates that it is the underlying X-ray emission mechanism in general that does not evolve with time.
Although we do not have a well developed physical model for the X--ray emission and variability in AGN, these implications put another observational constraint for any future attempts.

\begin{acknowledgements}
I.E.P. thanks the University of Naples Federico II for the financial support provided by the International Mobility program. M.P. acknowledges financial support received through the agreement ASI-INAF n.2017-14-H.O. W.N.B. acknowledges support from Chandra X-ray Center grant GO9-20099X and the V.M. Willaman Endowment. F.E.B. acknowledges support from ANID-Chile BASAL AFB-170002 and FB210003, FONDECYT Regular 1200495 and 1190818, and Millennium Science Initiative Program  – ICN12\_009. Y.Q.X. acknowledges support from NSFC grants (12025303 and 11890693), the K.C. Wong Education Foundation, and the National Key R\&D Program of China No. 2022YFF0503401. B.L. acknowledges financial support from the National Natural Science Foundation of China grant 11991053. D.D. acknowledges support from PON R\&I 2021, CUP E65F21002880003. This work has made use of lightcurves provided by the University of California, San Diego Center for Astrophysics and Space Sciences, X-ray Group (R.E. Rothschild, A.G. Markowitz, E.S. Rivers, and B.A. McKim), obtained at \url{https://cass.ucsd.edu/~rxteagn/}.
\end{acknowledgements}

\bibliographystyle{aa}
\bibliography{main}

\end{document}